\documentclass[11pt,conference]{IEEEtran}
\usepackage[document]{ragged2e}
\usepackage[english]{babel}
\usepackage{ragged2e}
\usepackage{graphicx}
\usepackage{float} 
\usepackage{fancyhdr}
\usepackage[colorlinks,citecolor=blue]{hyperref}
\usepackage{caption}
\usepackage{subcaption}
\fancypagestyle{firststyle}
{
\fancyhf{}
\lhead{\footnotesize{}}
\rhead{\footnotesize{May 2, 2022}}
\fancyfoot[C]{\thepage}

}
\begin{document}
\title{
Offloading Data Center Tax
}

\author{
    \IEEEauthorblockN{
    Akshay Revankar\IEEEauthorrefmark{1}, 
    Charan Renganathan\IEEEauthorrefmark{1},
    Sartaj Wariah\IEEEauthorrefmark{1}
    }
\IEEEauthorblockA
	{
        \begin{tabular}{cc}
            \begin{tabular}{@{}c@{}}
                    \IEEEauthorrefmark{1}Department of Electrical and Computer Engineering\\
            \end{tabular}
        \end{tabular}\\
                    akshayrevankar@cmu.edu, \{crengana,swariah\}@andrew.cmu.edu\\
                    Carnegie Mellon University\\
    }
}
\maketitle

\thispagestyle{firststyle}
\justify
\begin{abstract}
The data centers of today are running diverse workloads sharing many common lower level functions called tax components. Any optimization to any tax component will lead to performance improvements across the data center fleet. Typically, performance enhancements in tax components are achieved by offloading them to accelerators, however, it is not practical to offload every tax component. The goal of this paper is to identify opportunities to offload more than one tax component together. 
We focus on MongoDB which is a common microservice used in a large number of applications in the datacenter. We profile MongoDB running as part of the DeathStarBench benchmark suite, identifying its tax components and their microarchitectural implications. We make observations and suggestions based on the inferences made to offload a few of the tax components in this application.

\end{abstract}
\justify

\IEEEpeerreviewmaketitle

\section{Introduction}
Modern-day cloud platforms host a plethora of third-party web applications on their large ‘warehouse-scale’ data centers. Even with a large diversity in these web applications, we often find that many developers inherently use common underlying libraries and services for inter-instance networking, storage, compression, etc. \\

It is therefore critical to understand common patterns across diverse applications that contribute to significant part of the total cycles and thereby enable future optimizations across the hardware and software stack. Prior work \cite{wsc} has shown that nearly 25-30\% of the cycles pertaining to workloads running in the data center comprise common lower-level functions dubbed ``datacenter tax". \\

Although there’s significant evidence that many data center workloads share these tax components, many solutions target to offload them individually with custom hardware accelerators but there are a whole host of significant challenges (such as managing heterogeneity) in attempting this \\

This paper aims to pick up one of the most common back-end (micro-)service used in modern day applications for storage, MongoDB, and profile and characterize this workload at a micro-architectural level. We target a instance-level profiling at peak workload conditions using a well-known benchmark suite, the DeathStarBench \cite{dsb}. We present a quantitative analysis of the correlation between several data center tax components and highlight the micro-architectural trends seen during the execution of the functions associated with these tax components. \\

The paper suggests improvements that can be made in terms of reducing resource consumption, reducing stack overheads and offloading functions while keeping the 99th percentile latency within the existing margins. We provide a qualitative study and propose different paths that we can look into for benefiting two tax components, network and data allocation, together at datacenter level. \\

\textbf{Key Contributions:}
\begin{itemize}
    \item Identified correlations between the impacts of different tax components.
    \item Analyzed the microarchitectural impacts of tax components in MongoDB.
    \item Presented suggestions to offload the multiple components together.
\end{itemize}

\section{Background and Motivation}
The first step in reaping any kind of performance gain across the hardware or software stack involves a thorough characterization of workloads across varying load conditions. Additionally, by understanding certain bottlenecks and microarchitectural implications corresponding to an application, one can make well-informed decisions to potentially enhance its performance. Furthermore, the possibility of reduced resource allocation could potentially improve the performance on co-located applications.\\

Most existing studies \cite{prot,dagger} characterize only a single datacenter tax component. The focus of this study is to find opportunities to offload two or more of these components together. It's seen that optimizing for one bottleneck may have negative and diminishing effects on another component. Many existing solutions \cite{ftom,rob} attempt to reduce tail latency, however, this paper attempts to take a deeper dive into potential resource utilization improvements without negatively impacting the tail latency. \\

The rest of this section provides a background on the benchmark suites, the profiling tools, and the utilities used for our study and implementation.

\subsection{DeathStarBench}
DeathStarBench is an open-source benchmark suite that includes six end-to-end services representative of large warehouse-scale workloads. We are using one of the available microservice-based applications, the Social Network service.\\

\subsubsection{Social Network service}
\begin{figure}[h]
\includegraphics[width=0.5\textwidth]{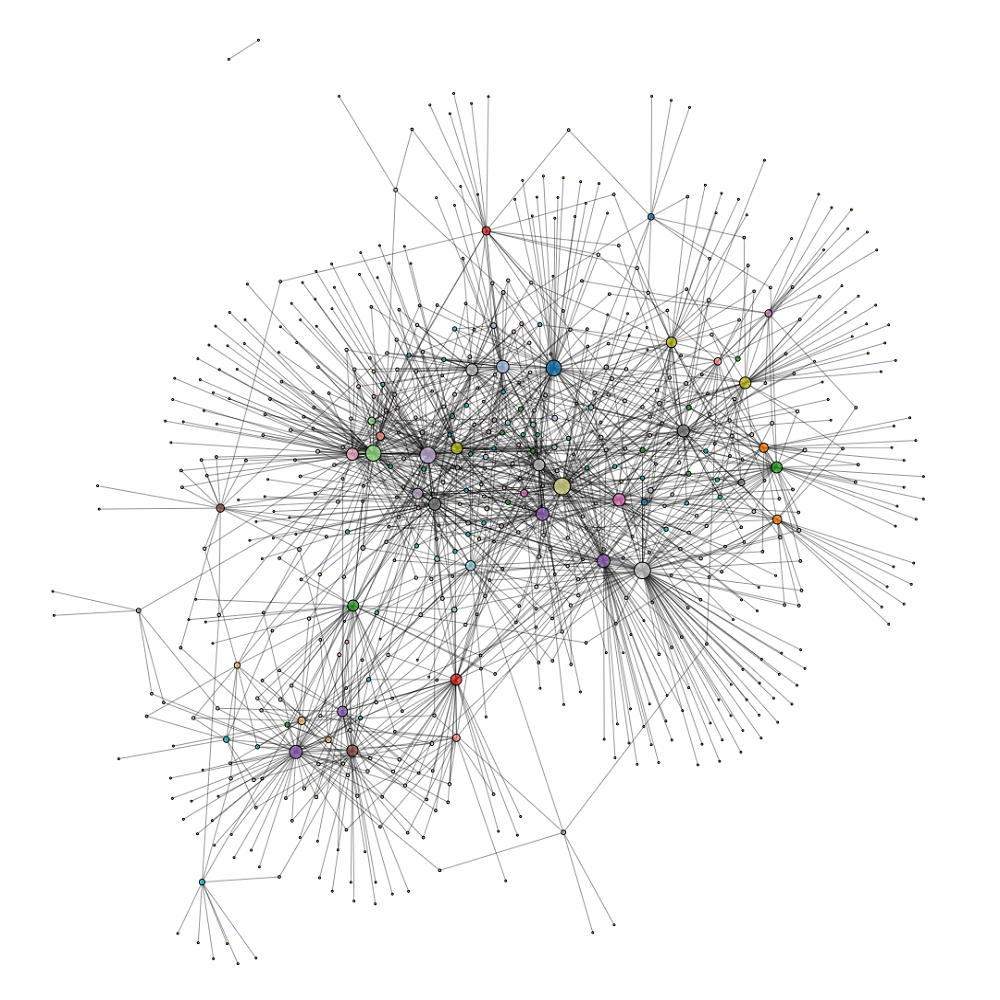}
\caption{A small social graph extracted from Facebook used in the study}
\label{fig:network_graph}
\end{figure}

The Social Network service consists of 36 microservices that perform various functions such as load balancing, machine learning for recommendations, caching, and persistent storage, communicating with each other via Thrift RPCs. Users can read, favorite, write, repost posts and send messages. Posts, composed of text, media and tags, are broadcast to all followers of a user. The service can be initialized and registered with different-sized social graphs \cite{reed98} similar to Figure \ref{fig:network_graph}. \\

The client requests pass through a load balancer implemented  with the Nginx web server, pass downstream to microservices responsible for various functions and reach leaf services consisting of 6 different MongoDB instances for persistent storage of posts, profiles, media, and recommendations.\\

\subsubsection{MongoDB}
MongoDB is a document-oriented NoSQL database program that uses BSON objects for storage. \cite{mongo} It is a widely used backend storage service for cloud applications. \\

In order to analyze MongoDB server behavior, mongod instances include a Full-Time diagnostic Data Collection (FTDC) that collects detailed information about operations run across a mongod instance, it is also used to recreate the database state in the occurrence of a failure. However, we disable it to reduce overhead and focus on the core functions. The mongod instances of DeathStarBench also use the WiredTiger storage engine (over in-memory) due to persistent storage requirements. \\

\subsubsection{Workload generator}
\begin{figure}[h]
\centering
\includegraphics[width=0.5\textwidth]{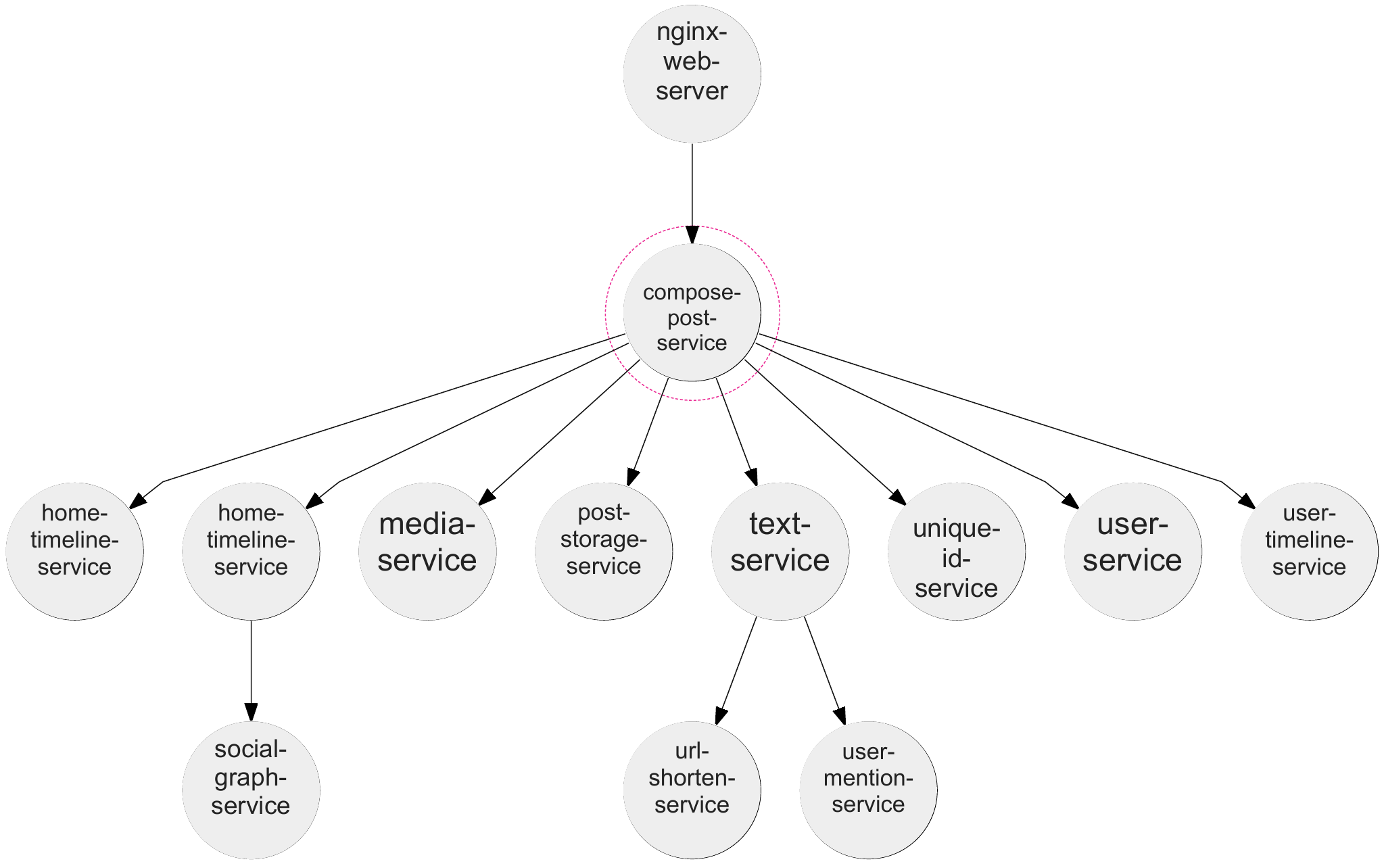}
\caption{The compose post workload flow}
\label{fig:compose-post}
\end{figure}

In the social network service, a user can compose posts, read home timelines and read user timelines. Each of these activities corresponds to a workload generator in the DeathStarBench suite. The workload generator allows the user to vary parameters such as the number of threads, the number of connections, duration, and requests per second. As a measure of performance, it provides the latency distribution for the workload scheduled. The compose post workload is “write-heavy” whereas the other workloads are dominated by other services like recommendation engines. The compose-post service triggers several downstream services as shown in Figure \ref{fig:compose-post}.\\

\subsubsection{Docker Swarm Coordination}
The default configuration for the social network service of the DeathStarBench provides a configuration for deploying the microservices as a Docker swarm, i.e. deploy one or more microservices distributed across multiple nodes at random.\\

However, we wanted to isolate the MongoDB service specifically onto a dedicated system in order to prevent interference between different microservices at the microarchitectural level. We hence used constraints within the Docker configuration and a 2-node setup to deploy MongoDB instances on one node and the rest of the microservices on another node.

\subsection{Performance Analysis}
\subsubsection{Perf}
Perf is a linux profiling utility used to instrument hardware performance counters, software performance counters, tracepoints, and dynamic probes. In this paper, we sample hardware counters related to the memory subsystem like CPU cycles, TLB, L1, LLC loads, stores and misses. Due to hardware limitations, perf can only sample a limited number of hardware counters ($\sim$4) at a time. If this limit is exceeded, it reverts to time multiplexing between the counters which results in less accurate measurements.\\

Perf supports branch tracing with Intel’s Last Branch Records (LBR), dwarf, or frame pointers. Among these, LBR is the most fine-grained; it saves executed branches in the Intel Special Branch Trace Store or Intel Processor Trace in newer systems. We used the generated call graph to associate a hardware metric to a function call. \\

\subsubsection{Flame Graph}
 Flame Graphs \cite{flamegraph} are used to visualize stack traces in order to determine the most frequently traversed code paths. The width of each frame determines how often it was present in the stacks. Flame graphs are generated by capturing the stacks through perf, folding them into single lines, and finally using these folded files to render SVGs. \\

The flame graphs clearly denote the ancestries and by scanning through these diligently, it’s possible to attribute each set of samples to a specific tax component. \\

\subsubsection{Breakdown of tax components}
We have binned different tax components based on the lowest hierarchy in the call stack. For example if network operation is the cause of memory allocation, we categorize this component as network instead of memory allocation.\\

\begin{itemize}
    \item \textbf{Network} - MongoDB uses the Boost Async I/O library for most network interactions
    \item \textbf{Compression} - MongoDB uses the Snappy fast data compression/decompression library for perform compression, mostly not falling within the network path
    \item \textbf{Memory Allocation} - All generic standard libraries and variants that perform memory allocation
\end{itemize}

\subsection{Intel\textregistered \  Resource Director Technology (RDT)}
Intel RDT provides capabilities for cache and memory allocation and monitoring exposed to Linux via the \texttt{resctrl} file system. \\

\subsubsection{Cache Allocation Technology (CAT)}
The Cache Allocation Technology \cite{CAT} allows software-level control on the amount and location of cache space in the last-level cache (LLC) that can be consumed by a given thread, app, VM or container of applications. Multi-tenant VMs are ubiquitous in today’s data center cloud setup and CAT helps provide isolation and dedicated cache resources ensuring consistent performance and prioritization of interactive applications by avoiding any performance effects due to resource conflicts.\\

\subsubsection{Memory Bandwidth Monitoring (MBM)}
The Memory Bandwidth Monitoring features \cite{MBM} of RDT help in collecting per-thread memory bandwidth monitoring for all threads. To understand an application behavior in detail, it is important to monitor the memory bandwidth. Some applications can have low cache sensitivity due to either too small or too large working sets that do not fit well in the cache and thereby under or over utilize memory bandwidth towards the main memory.\\

\section{Methodology}
\subsection{Experimental Setup}
We deploy the social network service of the DeathStarBench suite in Docker swarm mode and isolate all the MongoDB instances on a single machine to avoid potential interference. The SKU of systems used in the experimental setup is Intel\textregistered \ Xeon\textregistered \ CPU E5-2640 v4 @ 2.40GHz processors, which is a 10 core CPU per socket with 320 KiB L1d, 320 KiB L1i, 2.5 MiB L2, and 25 MiB L3 cache sizes. The last level cache has 20 ways (18 usable) and supports partitioning using Intel Cache Allocation Technology.\\

\subsection{Determining the load for further analysis}
\begin{figure}[h]
\centering
\includegraphics[width=0.5\textwidth]{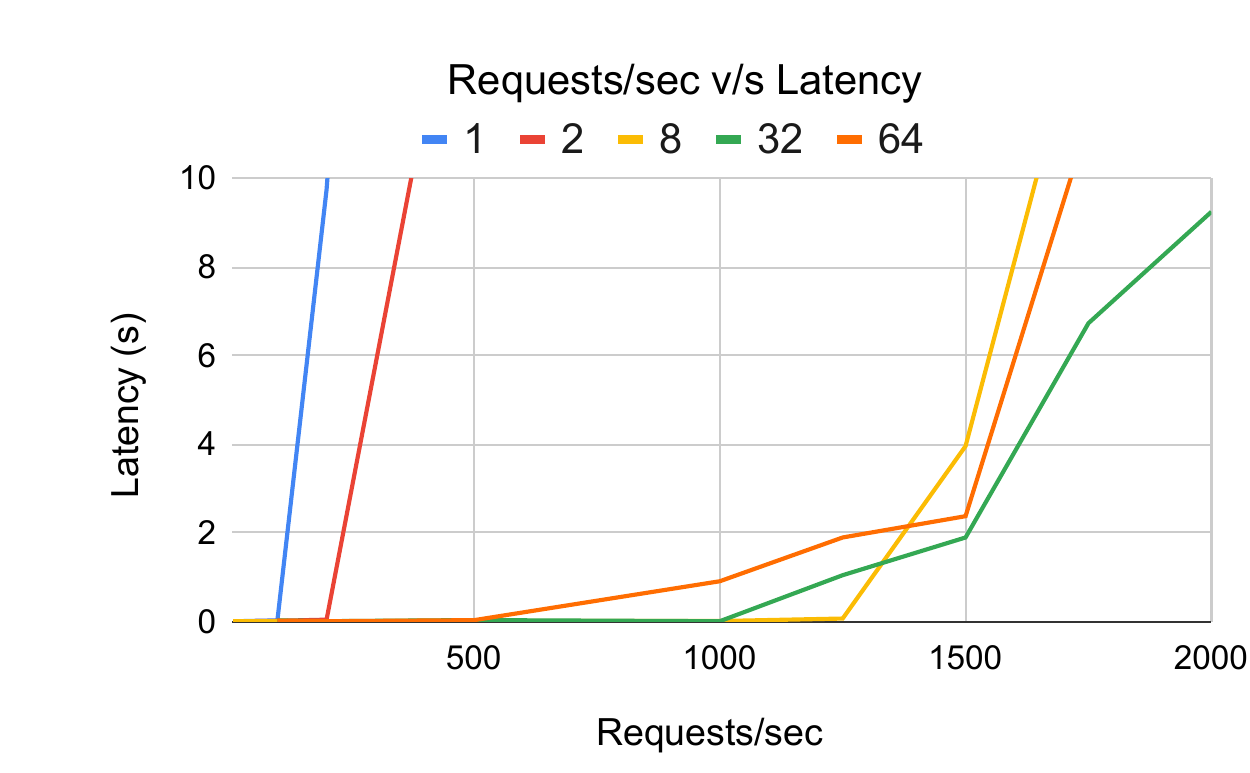}
\caption{Peak workload calculation for different Requests-per-Seconds at constant connection values}
\label{fig:latency_knee}
\end{figure}

In order to determine load conditions for the workload, we use the workload generator and fix the number of connections while sweeping through the requests per second (RPS) to track the variation in tail latency. The RPS rate at which the 99th percentile latency rises exponentially is used to determine the peak RPS (the knee in the graph). We observe the knee of the latency curve at 1750 RPS at 32 connections as seen in Figure. \ref{fig:latency_knee}.  For the remaining analysis, low load (10\%), medium load (30\%), and high load (70\%) will be used corresponding to RPS values of 175, 525, and 1225 respectively. The microarchitectural implications of the workload are measured against varying loads in the following sections. \\

\subsection{Workload Profiling}
Since the exploration space of our characterization was quite vast, we fixate upon MongoDB as a part of the compose-post service flow which is a part of the Social Network service. \\

We identify the process ID of the docker container running the MongoDB instance of interest and perform all our measurements on it using the perf tool. At each load, we sample four hardware metrics at a time at the default sampling frequency of 4000 Hz. First, all metrics (loads, load-misses, stores, store-misses) related to L1 cache are sampled, followed by LLC, and finally the TLB. We were not able to sample L2 metrics using perf as it is not supported on our system.\\

\begin{figure}[h]
\centering
\includegraphics[width=0.4\textwidth]{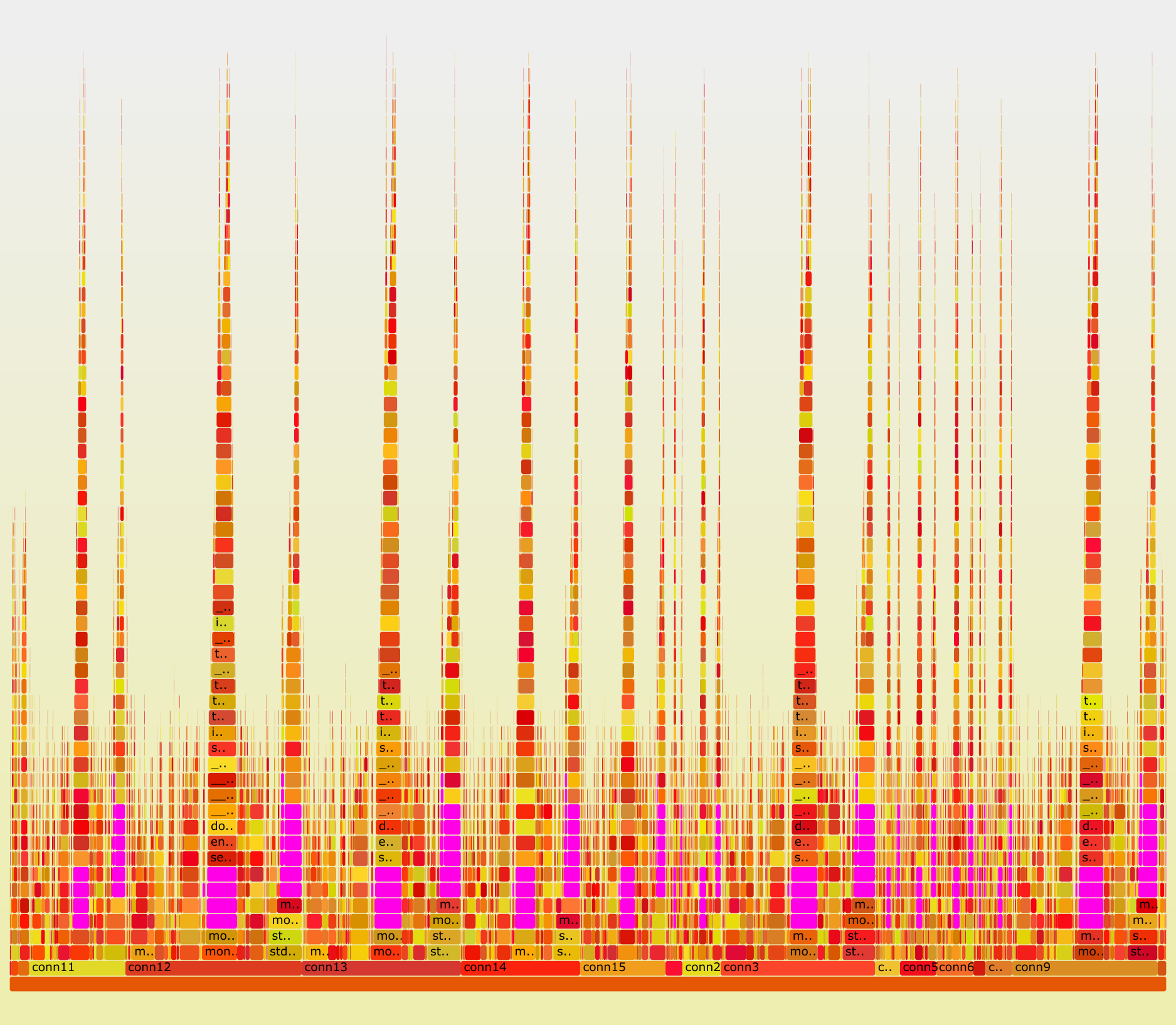}
\caption{Flamegraph showing stack traces on MongoDB for 1225 RPS load. The regions highlighted in magenta denote network operations}
\label{fig:flamegraph}
\end{figure}

The raw data from perf is processed to display the trace output. The trace output is parsed to generate the stack trace using the stack collapse programs. These folded stack traces can be visualized as flame graphs as shown in Figure \ref{fig:flamegraph} for quick analysis and parsed to extract call stack composition.\\

\begin{table*}[t]
    \centering
    \begin{tabular}{lllllll}
        \hline
        Load (RPS) & Network & Memory Allocation & Compression & Data Movement & Hashing & Serialization \\ \hline
        175        & 21.97   & 5.05      & 1.70        & 0.54          & 0.59    & 0.19          \\
        525        & 26.32   & 5.62      & 2.05        & 0.53          & 0.55    & 0.23          \\
        1225       & 25.99   & 6.20      & 2.28        & 0.64          & 0.57    & 0.23          \\ \hline
    \end{tabular}
    \caption{Contribution percentage of different tax components for varying Requests/sec load}
    \label{table:cycle_perc}
\end{table*}

To identify the time spent performing tax component functions, we go through all the unique functions called by the process and bin them in a category suggested by their ancestral hierarchy and library functions. Once the functions associated with a particular tax component are identified, we parse the trace to find the total CPU cycles consumed by that tax component. The same technique is used to find the other hardware metrics related to the tax components.\\

\begin{figure}[h]
\centering
\includegraphics[width=0.58\textwidth]{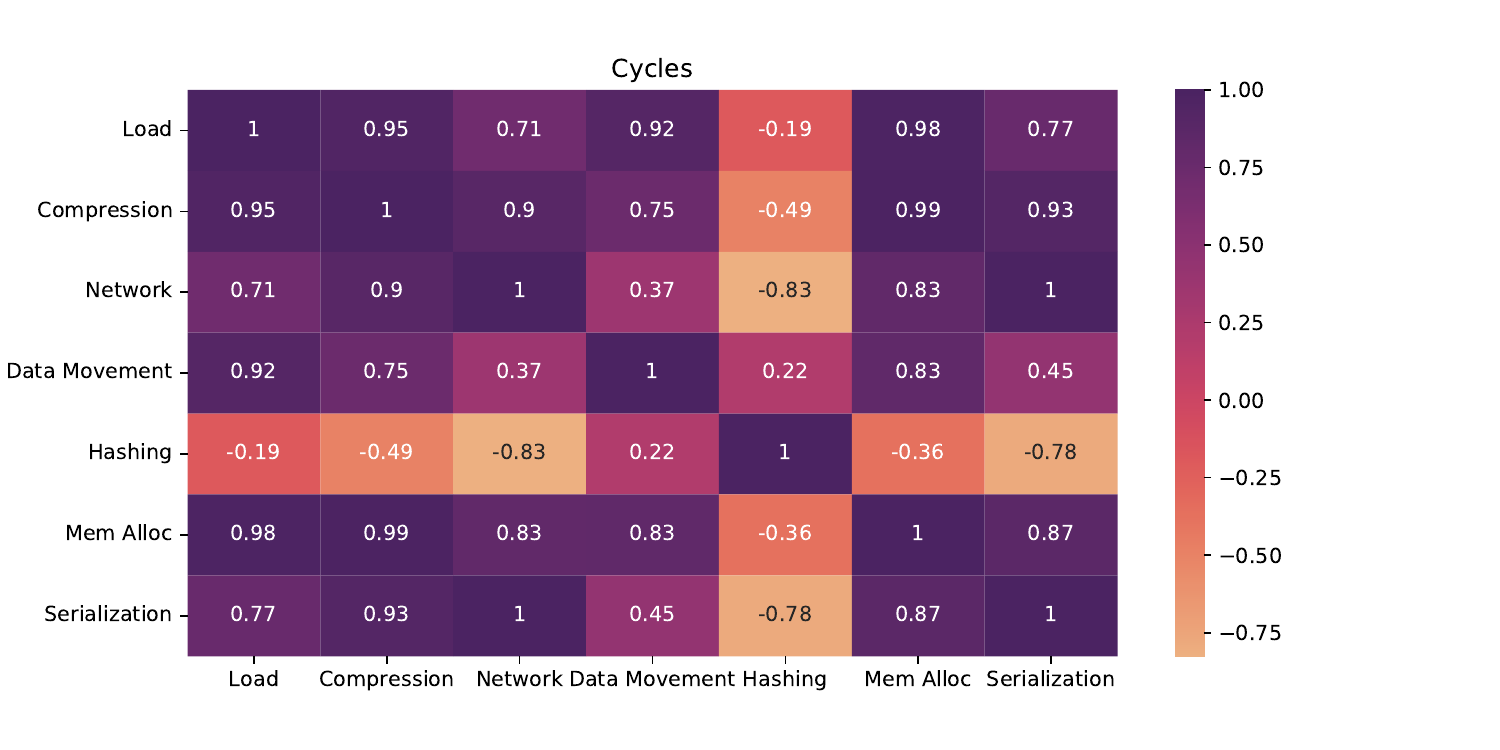}
\caption{Correlation between cycles spent running each tax component}
\label{fig:corr:cycles}
\end{figure}

This allows us to find the relationship between the application load and the hardware metrics. We find the change in the number of samples and the percentage of total samples at each load to find the cross correlation between the hardware metrics, which are visualized in Figure \ref{fig:corr:cycles}.

\section{Evaluation and Insights}

There are two key considerations in determining which tax components can be potentially offloaded. Firstly, the two components should have a significant contribution in terms of samples, and additionally, their correlation should remain consistently high (greater than 0.95) across all metrics corresponding to a specific memory structure. \\

Amongst the 6 tax components, we note that the significant contributors towards cycle count are network, compression, and memory allocation as seen in Table \ref{table:cycle_perc} , whereas, the other tax components' contributions were negligible. \\ 

Within these select 3 components, further analysis is performed by evaluating their correlations across different load conditions. From Figure \ref{fig:corr:cycles}, we observe that memory allocation and compression are strongly correlated ($>0.95$). It also shows that network and compression are weakly related, which is a counter intuitive result as these functions should logically follow each other. This might be an artifact of MongoDB and may not hold across other applications that also use network and compression operations. Thus, due to our limited knowledge regarding true application behavior, we generate correlation matrices across all 6 components. \\

\begin{figure}[h]
     \centering
    \begin{subfigure}[b]{0.4\textwidth}
         \centering
        \includegraphics[width=\textwidth]{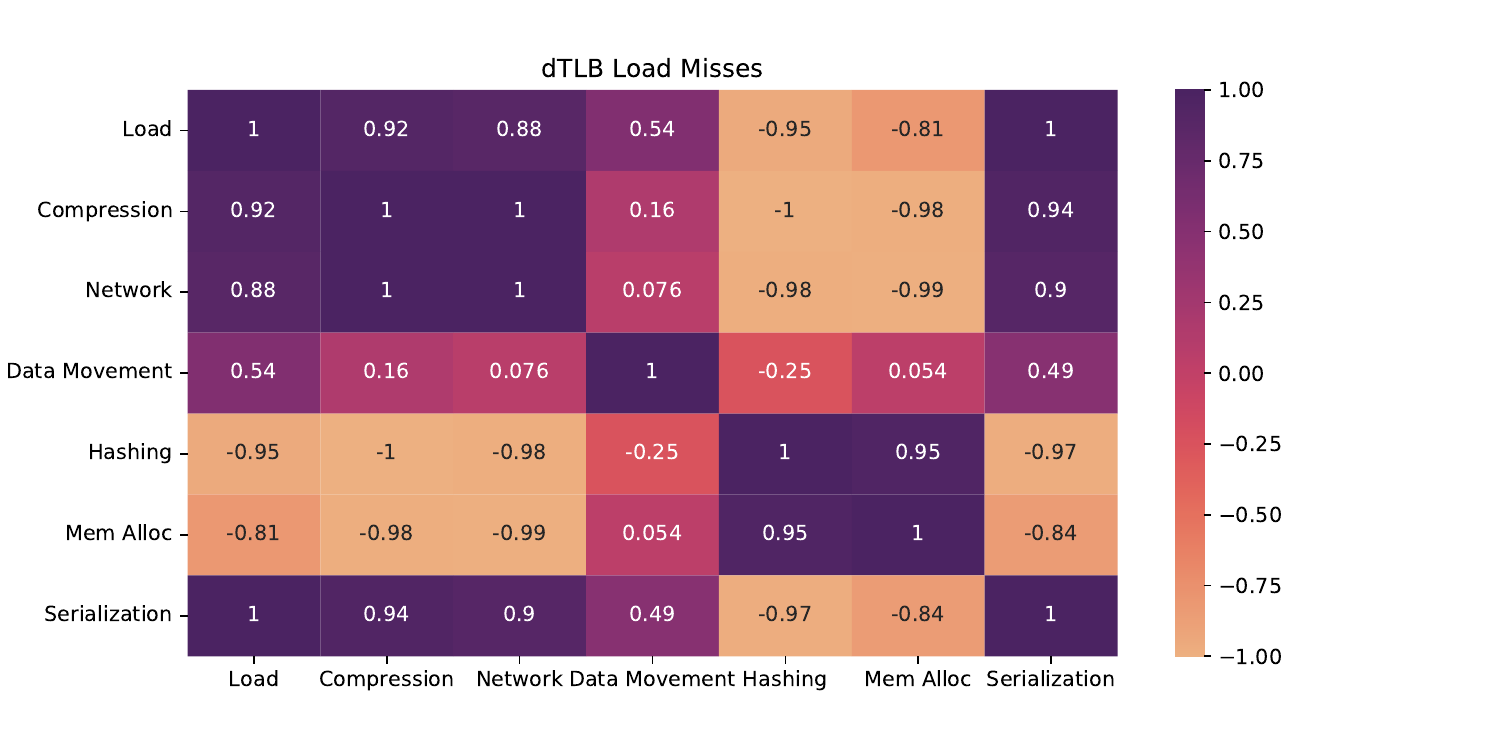}
        \caption{dTLB load misses}
        \label{fig:corr:dtlb_load_misses}
    \end{subfigure}
    \begin{subfigure}[b]{0.4\textwidth}
         \centering
        \includegraphics[width=\textwidth]{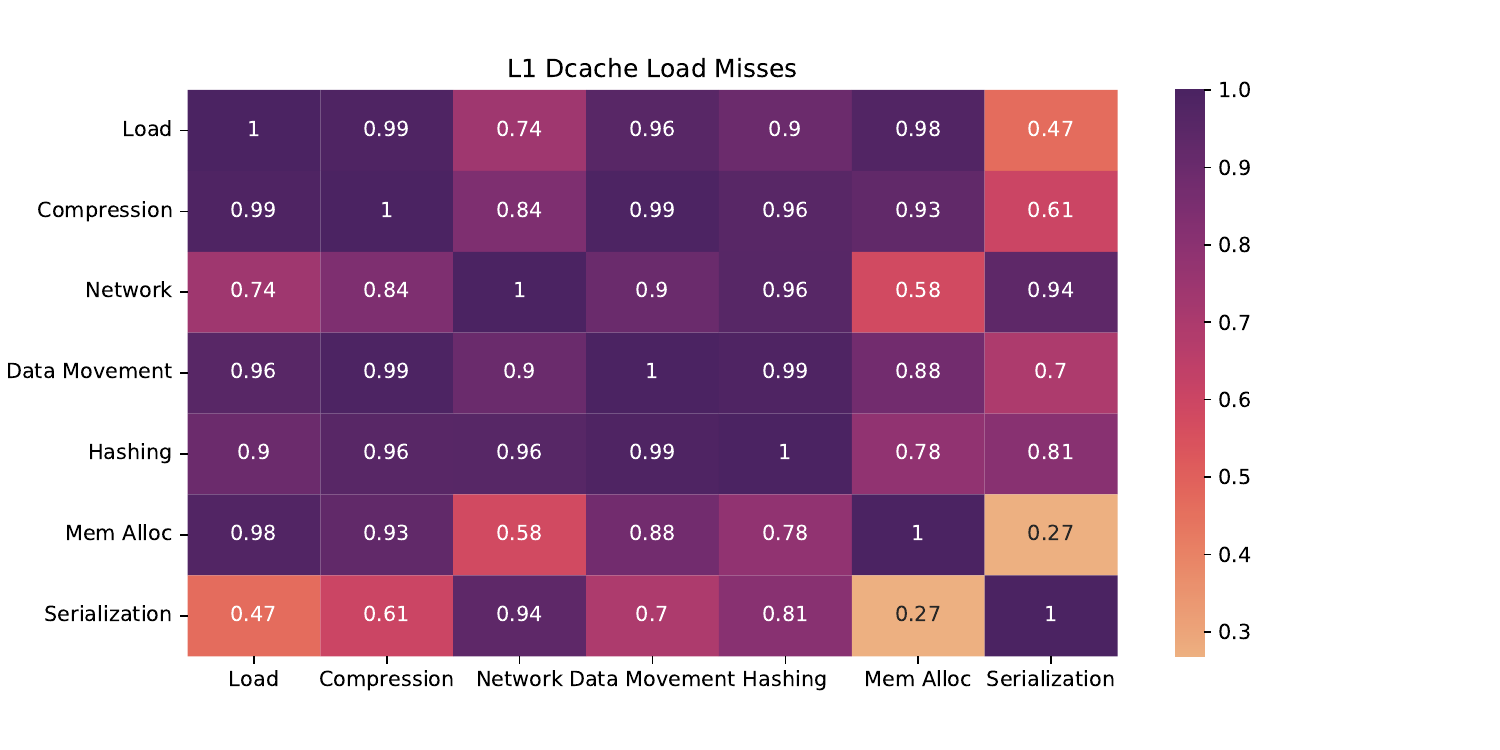}
        \caption{L1 Dcache load misses}
        \label{fig:corr:l1_dcache_load_misses}
    \end{subfigure}
    \begin{subfigure}[b]{0.4\textwidth}
         \centering
        \includegraphics[width=\textwidth]{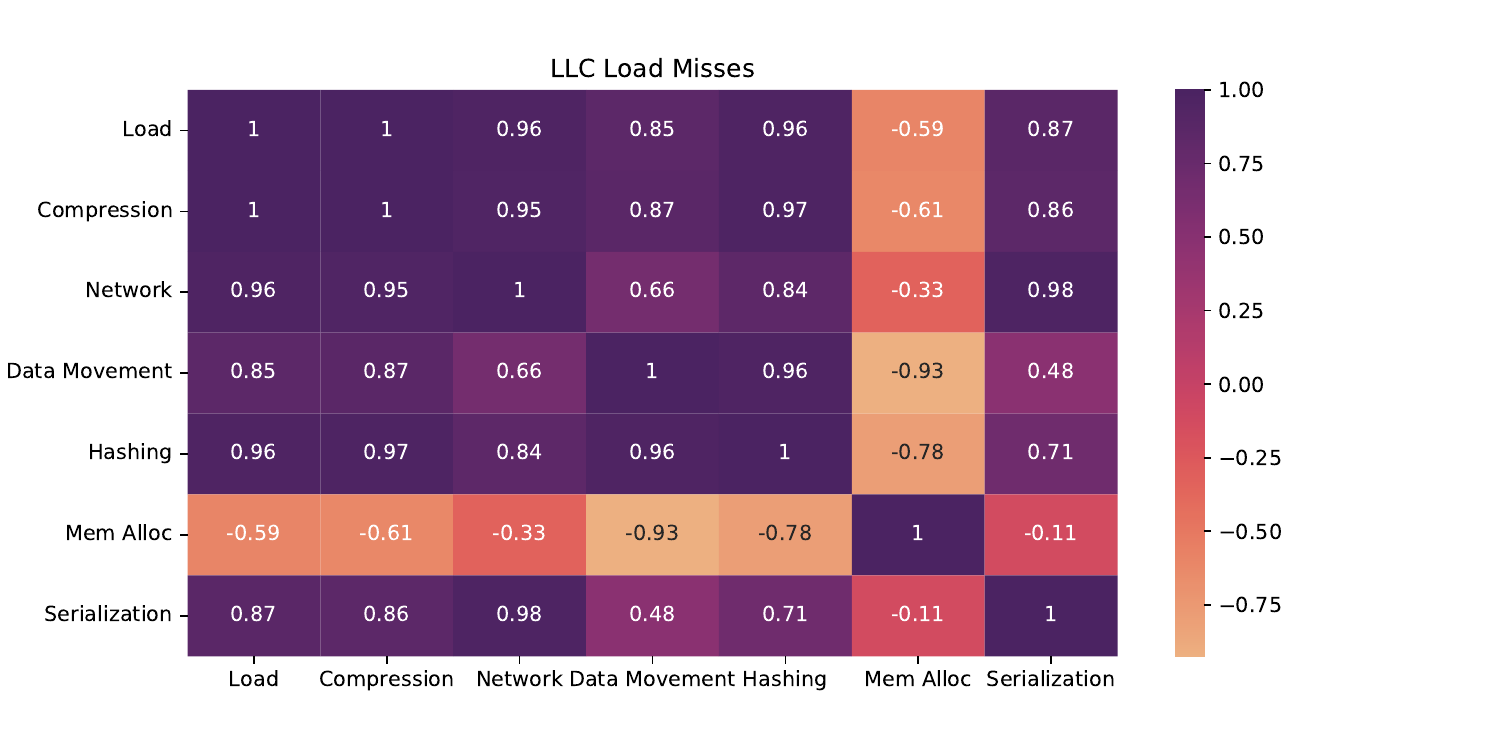}
        \caption{LLC load misses}
        \label{fig:corr:llc_load_misses}
    \end{subfigure}
    \caption{Correlation between metrics cycles spent running each tax component}
    \label{fig:corr:dtlb_l1_llc}
\end{figure}

\begin{table*}[t]
    \centering
    \begin{tabular} {l|ccc|ccc|ccc}
    \hline
        ~ & \multicolumn{3}{c|}{Network} & \multicolumn{3}{c|}{Compression} & \multicolumn{3}{c}{Memory Allocation} \\ 
        ~ & Loads & Load Misses & Stores & Loads & Load Misses & Stores & Loads & Load Misses & Stores \\ \hline
        Low & 30.82 & 19.28 & 24.17 & 0.34 & 0.35 & 8.99 & 6.02 & 13.06 & 8.89 \\
        Medium & 31.60 & 26.20 & 24.82 & 0.45 & 0.42 & 8.61 & 5.59 & 13.54 & 8.79 \\
        High & 27.89 & 30.94 & 25.04 & 0.28 & 0.56 & 8.36 & 5.77 & 12.70 & 7.72 \\ \hline
    \end{tabular}
    \caption{LLC events for varying requests/second loads with respect to network, compression and memory allocation}
    \label{table:llc_table}
\end{table*}

In Figure \ref{fig:corr:l1_dcache_load_misses}, we observe that memory allocation and compression are significant contributors for all L1 events and are strongly correlated with each other. Hence, any optimization in L1 cache performance will significantly benefit both components. However, it is non-trivial to increase L1 capacity as it may increase access latency and area requirements. \\

By observing Figure \ref{fig:corr:llc_load_misses} in addition to the other correlation matrices(not shown) generated for LLC events, we find that memory allocation, network, and compression remain relatively consistent across the metrics under the varying load conditions as shown in Table \ref{table:llc_table}. This implies that these components are not bound by LLC performance and some resources allocated to the cache can be reclaimed. \\

\begin{figure}[h]
\centering
    \includegraphics[width=0.5\textwidth]{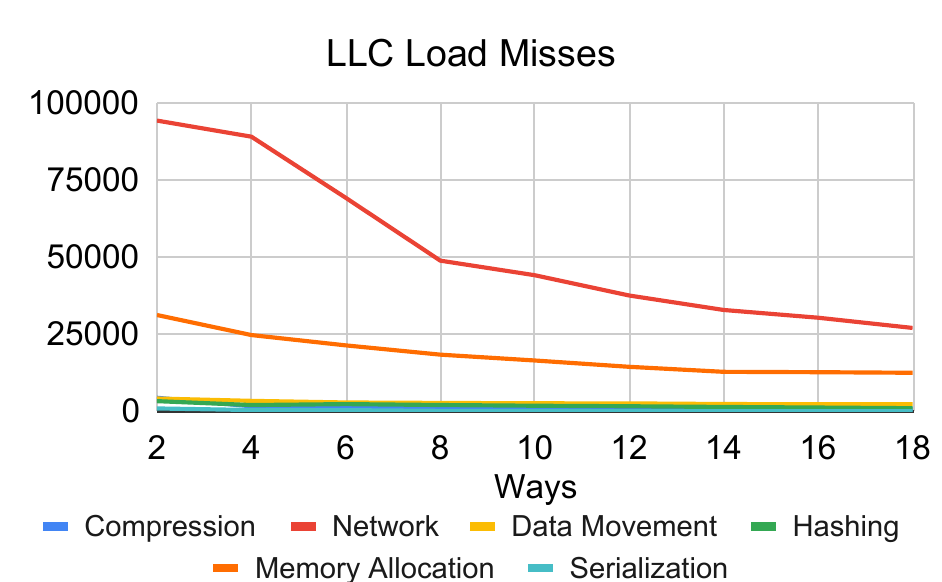}
    \caption{Impact on LLC load-misses w.r.t Number of ways v/s overall cycles equivalent}
    \label{fig:llc_ways}
\end{figure}

\begin{figure}[h]
     \centering
    
    \begin{subfigure}[b]{0.45\textwidth}
    \includegraphics[width=\textwidth]{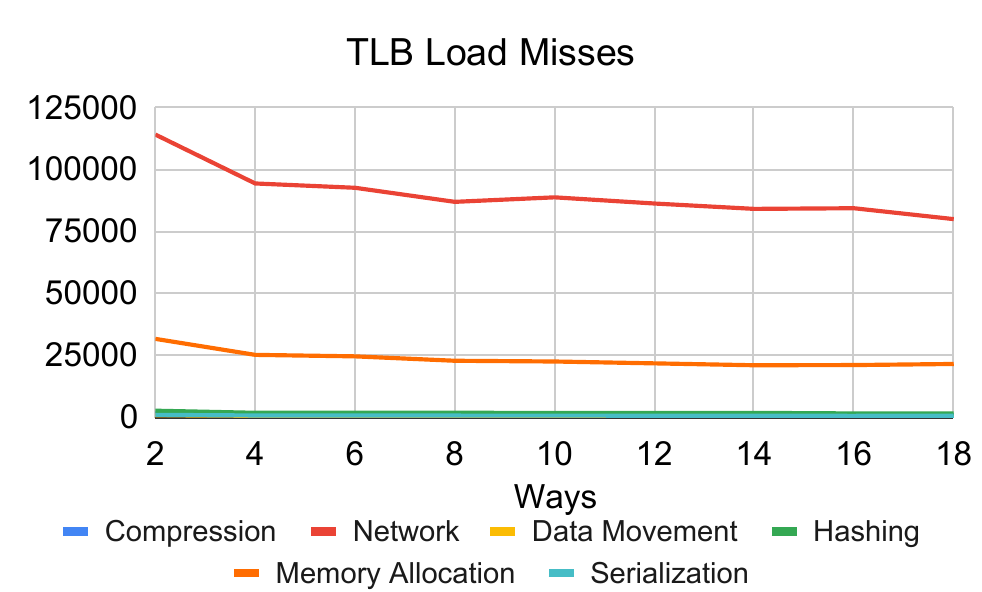}
    \caption{TLB load-misses}
    \label{fig:tlb_ways}
    \end{subfigure}
    
    \begin{subfigure}[b]{0.45\textwidth}
    \includegraphics[width=\textwidth]{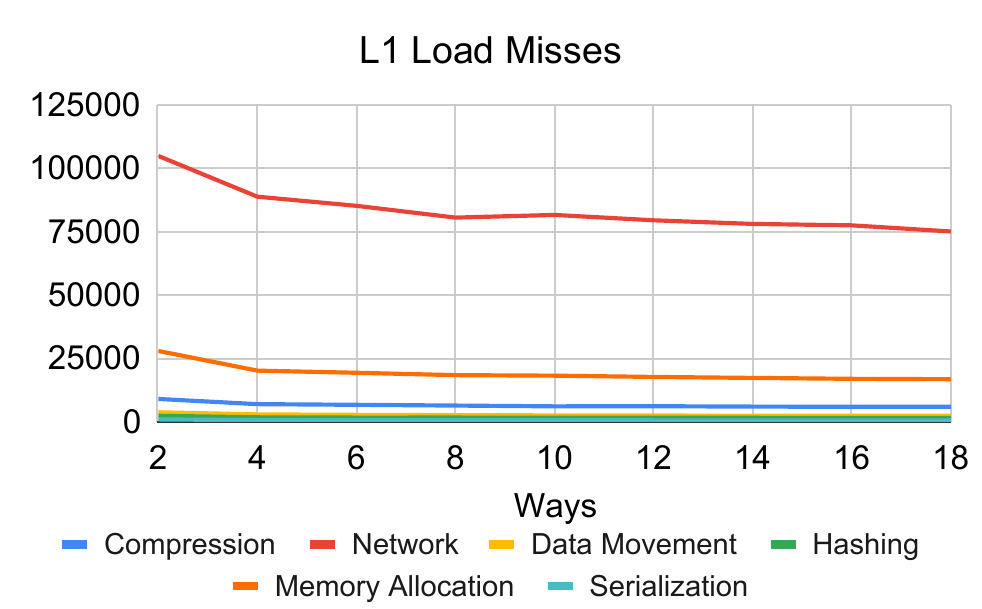}
    \caption{L1 load-misses}
    \label{fig:l1_ways}
    \end{subfigure}
    \caption{Impact w.r.t Number of ways v/s overall cycles equivalent}
    \label{fig:l1_tlb_ways}
\end{figure}

Armed with this realization, we attempt to find the implications of limiting the LLC cache capacity by restricting the number of ways through cache allocation technology (CAT). Figure \ref{fig:llc_ways}  shows us that additional cache capacity beyond 8 ways yields diminishing returns in reducing LLC load-misses and remains stable across the other memory structures as seen in Figure \ref{fig:l1_tlb_ways}. This leads us to believe that this additional area may be repurposed in resizing other components without negatively effecting the upstream structures. \\

\begin{figure}[h]
\centering
    \includegraphics[width=0.5\textwidth]{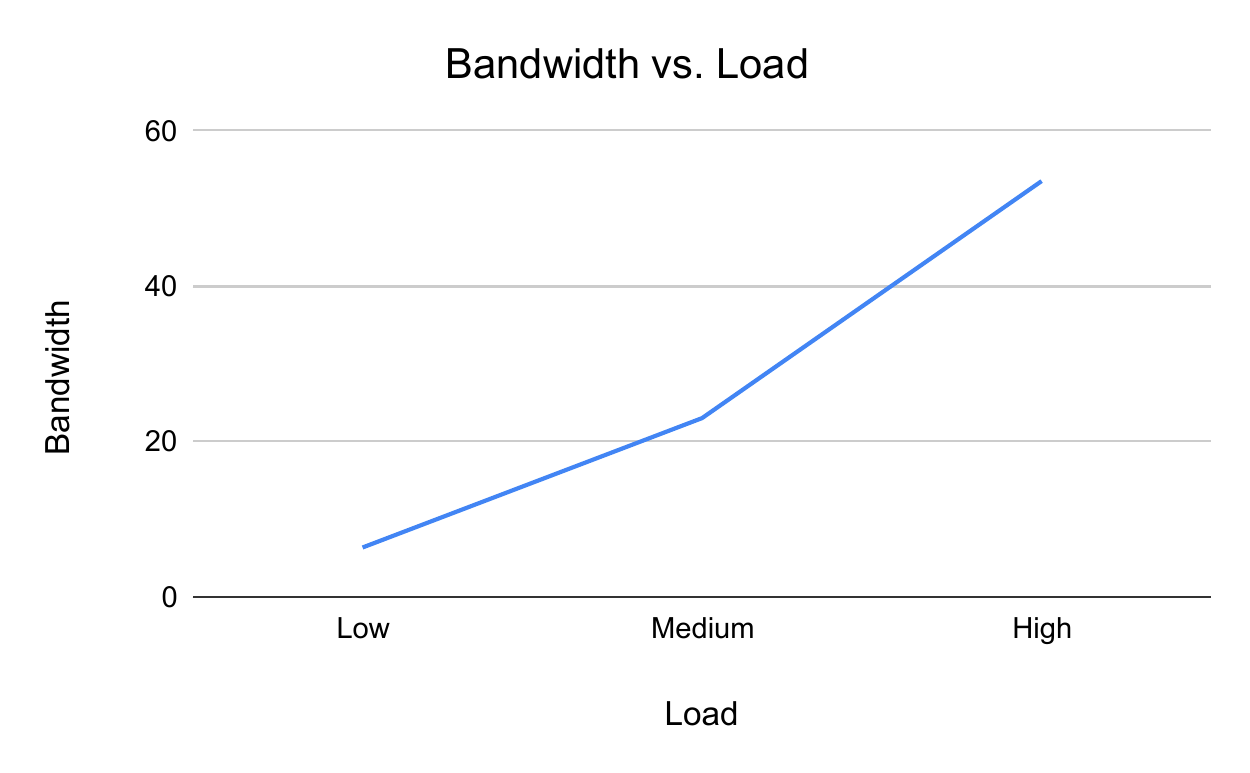}
    \caption{Memory Bandwidth Utilization for three load conditions}
    \label{fig:mbm}
\end{figure}

Additionally, we monitored the memory bandwidth utilization at the different loads using Intel MBM tool. The utilization observed is as expected; it increases with the load as seen in Figure \ref{fig:mbm}. This suggests that the application is not memory bandwidth bound.\\

From these insights, we propose reducing the resource area dedicated to the LLC. This could be used to increase the L1 size or build an on-chip accelerator. However, further work must be done to find the optimal L1 size and the application to accelerate. Network could possibly be a good contender as it had the highest impact.\\

These insights can be exploited without hardware changes. The network overhead can be reduced by using libOSes. The benefit of these kernel bypass systems like Demikernel \cite{demikernel} could be more widely applied if they were implemented into common asynchronous I/O libraries (like the C++ Boost library used within MongoDB). We could also offload network and memory allocation using existing NICs like Intel's with DPDK technology \cite{dpdk}, which is already widely deployed in modern data centers. //

\section{Related Work}
Although the inception of data center tax (``Profiling a warehouse-scale computer" \cite{wsc}) was in 2015, few prior studies seek to explore their combined implications. These studies suggest a wide range of investigation directions revolving around the usage of datacenter-specific SoCs to combat tax, cache prefetchers/cache partitioning to combat growing instruction footprints, and the potential trade-off of memory bandwidth for cores in low-bandwidth utilization cases. \\

``Accelerometer: Understanding Acceleration Opportunities for Data Center Overheads at Hyperscale" \cite{acc} build an analytical model to understand the opportunities for hardware acceleration. It finds the possible benefits of the acceleration strategy and threading model. \\

``Server engineering insights for large-scale online services" \cite{seilsos} discusses the dilemma data center providers face in the context of provisioning for infrequent peak load cases, as optimizing for such scenarios involves cost vs performance considerations. This leads us to believe that potential application-specific performance benefits may not be welcomed with open arms. \\

``Memory hierarchy for web search" \cite{mhws} extensively profiles web search and further showcases that the L3 caches are over provisioned for web search. They evaluate trading off L3 caches for the L4 as well as processing cores, to gain performance. Similarly, our work demonstrates the potential of using workload-specific SoCs in future data centers. \\

``A Hardware Accelerator for Protocol Buffers" \cite{prot}
was capable of profiling both serialization and de-serialization in an end-to-end manner and proposing a fully open-source hardware accelerator for protobufs. A key takeaway in such an implementation was the possibility of reusing their hardware blocks to accelerate other protobuf operations apart from serialization and deserialization. This lends us insight into the benefit of designing for flexibility in order to accommodate other non-trivial components. \\

Furthermore, \texttt{Dagger} \cite{dagger} was successful in incorporating reconfigurability in their system to offload network operations off-chip. Similar setups could potentially be a more reasonable approach towards incorporating accelerators in the data center.\\

\section{Future Work}
The manual analysis of applications is labor intensive. We assume that there exists the possibility of creating a standard methodology to profile and subsequently sub-categorize microservices based on similar metrics showcased in this paper. This may enable datacenter administrators to apply profile-guided optimizations. Additionally, for applications that exhibit erratic characteristics across loads, we believe that it's possible to deploy a light-weight tool with negligible overhead like Google Wide Profiling \cite{GWP} to make performance enhancing decisions on the fly. \\

Another key aspect we had looked over was the frequency of function calls and their corresponding contribution to the flame graphs. An ideal candidate for hardware acceleration would be a function called infrequently and consumes a large amount of cycles because the time saved by offloading the task will not be overshadowed by the latency of communicating with the accelerator. \\

Understanding the microarchitectural bottlenecks of microservices, enables us to make smarter application binning decisions. Once possible method may be to avoid co-locating microservices that stress the same microarchitectural structure and cause negative interference. Inversely, microservices that stress different structures would be good candidates to co-locate. \\

Unfortunately due to the limitations of our hardware, we were not able to sample L2 cache events. We would like to rerun our analysis on different hardware platforms to capture this information as well. This may or may not also confirm whether our insights are hardware agnostic. \\

In this study, we only focused on the tax components that had the impact. However, even an improvement of a fraction of a percent has large implications in warehouse scale computers. Thus, we could attempt to understand if tax components with low contribution still make sense to offload together. \\

A followup study into the feasibility of increasing L1 cache size will allow us to better understand the advantage of reclaiming the LLC capacity as suggested in this study. An initial analysis may be done with done with simulators running on FPGAs (as Intel CAT or OS based page colouring can't be used to partition the L1 cache). Such an exploration will find the optimal L1 size and will help decide the trade off between using the area for L1, LLC or an on-chip accelerator. \\

\section{Conclusion}
In order to better understand MongoDB as a component used across many large-scale data center services, we profiled MongoDB as a part of the DeathStarBench benchmark. In this paper, we showcased certain trends across MongoDB, its bottlenecks, and tax components. Using this we analyzed their correlation as well as the memory and I/O-based architectural metrics. Through our initial characterization, we further performed a study through the use of CAT to evaluate cache area trade-offs. \\

Our observations motivate several future directions across the hardware and software stack. The first step in supporting diverse workloads revolves around fully characterizing these workloads, and gaining insights on their microarchitectural implications. Furthermore, such characterization could prove to be beneficial for the provider, as well as the end-user in terms of both cost and performance. Additionally, modern-day datacenters need to be able to support an ever-growing application pool, and service an increasing number of users, paving way towards 

\section{Acknowledgements}
We would like to acknowledge Akshitha Sriraman for her constructive feedback. We would also like to thank CloudLab for providing us with the compute resources to perform our experiments for the study.


\end{document}